\documentclass[journal]{IEEEtran}
\usepackage{blindtext}
\usepackage{graphicx}
\usepackage{float}
\usepackage{amsmath}
\usepackage{amssymb}
\usepackage{comment}
\usepackage{url}
\usepackage{empheq}

\begin{document}

\title{Centralized versus Decentralized\\
Infrastructure Networks}

\author{
Paul D.~H.~Hines,~\IEEEmembership{Senior Member,~IEEE},
Seth Blumsack,~\IEEEmembership{Member,~IEEE},
and~Markus Schl\"apfer%
\thanks{P.H. was supported by NSF awards ECCS-1254549 and DGE-1144388, and DTRA award HDTRA110-1-0088. S.B. acknowledges support from NSF award CNS-1331761. M.S. was supported by the Army Research Office Minerva Programme (grant no. W911NF-12-1-0097).}
\thanks{P.~Hines is with the University of Vermont, School of Engineering and Complex Systems Center, Burlington, USA, paul.hines@uvm.edu}%
\thanks{S.~Blumsack is with Penn State University, University Park PA, USA, sethb@psu.edu}
\thanks{M.~Schl\"apfer is with the Santa Fe Institute, Santa Fe, USA, schlaepfer@santafe.edu}%
}

\maketitle

\begin{abstract}
While many large infrastructure networks,
such as power, water, and natural gas systems,
have similar physical properties governing flows,
these 
systems
tend to 
have distinctly different sizes and topological structures. 
This paper seeks to understand how these different size-scales and topological features can emerge from relatively simple design principles. 
Specifically, we seek to describe the conditions under which it is optimal to build decentralized network infrastructures, such as a microgrid,
rather than centralized ones, such as a large high-voltage power system.
While our method is simple 
it is useful in explaining why sometimes, but not always, it is economical to build large, interconnected networks 
and in other cases it is preferable to use smaller, distributed systems.
The results indicate that there is not a single set of infrastructure cost conditions under which optimally-designed networks will have highly centralized 
architectures. 
Instead, as costs increase we find that average network sizes increase gradually according to a power-law.
When we consider the reliability costs, however, we do observe a transition point at which optimally designed networks become more centralized with larger geographic scope. 
As the losses associated with node and edge failures become more costly, this transition becomes more sudden.
\end{abstract}

\section{Introduction}

One of the key goals of the United States and its allies during the United States/NATO conflict in Afghanistan was the improvement of infrastructure, particularly electricity infrastructure, in order to build good will among the Afghan people. 
One of these major infrastructure projects was the upgrade of the Kajaki Hydroelectric plant from 33 to 51 MW, a project designed to bring additional power to Kandahar, about 80 km to the southeast. 
However, moving the necessary heavy equipment from Kandahar to Kajaki through hostile territory proved to be one of the most difficult and costly operations of the Afghanistan conflict~\cite{zorpette2011re}.
And in the end (at least as of late 2015) the project was never completed because it proved impossible to move the concrete necessary to complete the project. 
At around the same time, the U.S. Agency for International Development (USAID) embarked on a less ambitious project to install several smaller (10 MW) diesel power plants in the outskirts of Kandahar. These plants are more expensive to operate given that they require diesel fuel, but because they were built close to the city, they were not nearly as complicated to install and continue to provide relatively reliable power for residents in that portion of the city~\cite{zorpette2011re}.
While there are many factors that contributed to the demise of the Kajaki Hydroelectric plant and the relative success of the more distributed diesel plants, 
it seems reasonable to ask whether a solution that is less reliant on long-distance transmission is, under these particular conditions, fundamentally better.
More generally, one might surmise that there exist general conditions under which decentralized solutions are fundamentally better. 
But what are those conditions?

Consider a second example: water distribution networks. 
In the United States alone there are more than 150,000 public drinking water networks that serve at least 25 people~\cite{EPA:2007water}.
In contrast, there are only 3 power networks: the Eastern, Western, and Texas interconnections.
Why is it that in the case of drinking water, smaller, more decentralized networks seem to be optimal, whereas for electric power, larger systems that span continents seem to be preferable?
Both systems have similar physical properties that govern flows.
Both systems transport largely interchangeable goods: one electron is as good as another, just as one water molecule is as good as another (given appropriate standards for cleanliness). 
However, these two systems have fundamentally different size scales.
A drinking water system serves, on average, 2000 people.
A power network serves, on average, 100,000,000 people.

Motivated by these, and many other, examples, 
this paper seeks to identify conditions under which it may be preferable to build and maintain large, centralized, interconnected infrastructure networks, 
versus constructing smaller, decentralized networks that effectively operate independently from one another. 

Large infrastructure systems are designed to deliver services in a way that balances a variety of potentially conflicting objectives including economic cost, environmental impact and reliability. These fundamental trade-offs become particularly challenging when societies face the potential for rapid infrastructure transitions.
This paper is motivated by two such distinct global infrastructure transitions.

The first transition is the growing push toward decentralized electric energy systems in more developed countries. 
The energy infrastructures 
in most industrialized countries have evolved into complex network structures \cite{Cotilla-Sanchez:2012}. 
However the growing movement toward the use of microgrids is a push back toward small, relatively independent systems~\cite{Lasseter:2011procIEEE}.
The historical case for large interconnected systems (economies of scale in generation and transmission) and increased redundancy through interconnection is being challenged by falling costs for distributed power generation~\cite{NETL:2013}, 
increased interest in smart micro-grids, 
and the insight that distributed systems may offer improved local reliability in some cases~\cite{Zerriffi:2007}.

The second transition is the rapid growth of infrastructure, including electric power, in less developed countries~\cite{Zvoleff:2009}. 
Prior work has argued that for political, geographic and economic reasons, the greenfield build-out of highly interconnected electric power infrastructure may not be desirable in developing-nation contexts, particularly in locations that are subject to elevated levels of stress~\cite{Zerriffi:2011}. 

While the contexts for infrastructure decisions in more-developed and less-developed nations differ, 
the basic question remains the same---%
given the need to build out power generation and delivery systems either incrementally (as in more-developed countries) or as a greenfield project (closer to the case for many less-developed nations), what mix of small-scale versus large-scale system architectures will best balance cost and reliability goals? Under what assumptions about cost, reliability and other factors would it be more advantageous to make either incremental or greenfield investments in decentralized system architecture?
This paper seeks to understand the conditions under which the transformation of large-scale systems into multiple smaller-scale systems, or the greenfield construction of multiple smaller-scale systems to serve a large geographic area, would yield improvements in cost and other measures of performance.

Our work is fundamentally concerned with the optimal planning of networks that deliver services or otherwise provide connectivity over physical space. As such, we model a single planner making optimal resource decisions, as distinct from game-theoretic approaches of network generation or the literature in random generation of synthetic networks ~\cite{Fabrikant:2003,Wang:2010}. 
While this topic has been of interest to geographers since the 1960s~\cite{Kansky:1960,Haggett:1969}, 
spatial network design has emerged only more recently as an area of 
scientific research~\cite{Gastner:2006}. 
Many applications of spatial network analysis focus on traffic or transportation networks~\cite{Garrison:1960,Gastner:2006,Kurant:2006,Chan:2011,Wilkinson:2012} 
or physical infrastructures that deliver information, such as the internet or mobile telephony~\cite{Gastner:2006,Lambiotte:2008}.
Others find that the constraints of geographic space can dramatically change the implications found in abstract network models~\cite{Barthelemy2011,Bashan:2013}.

Research on the spatial aspects of network design or performance, as distinct from data-driven empirical investigations of spatial network structure, has largely focused in two areas. The first is how the cost of adding edges or otherwise connecting nodes in space influences network structure and design choices \cite{Gastner:2006}, \cite{Barthelemy:2011}. The second strand utilizes known or theoretical spatial properties of networks to understand their performance in the case of attacks, failures or other contingency events \cite{Zerriffi:2007}, \cite{Wilkinson:2012}, \cite{McAndrew:2015}.

We build on this extensive body of work, and add to its relevance for electrical networks, in two ways. The first is to embed some most salient properties of electric power networks (namely Kirchhoff's Current Law)
into the type of cost-driven spatial network design problem discussed in \cite{Gastner:2006} and \cite{Barthelemy:2011}. 
These properties are important for electric power networks specifically because while expansion costs may be straightforward to parametrize in terms of spatial distance, actual network flows are not so simply represented. 
Further, production costs in real electrical networks are heterogeneous by technology and in space, because of regional variations in resource endowments and technology choices. 
The second is to consider a design objective that incorporates the 
costs of network operation,
network expansion,
and network (un)reliability.
While joint planning and operations models have been devised for incremental expansion decisions \cite{Vila:2010,Borraz:2016}, and for optimal topology control applications (e.g., \cite{Fisher:2008,Hedman:2010}),  our approach is different in its consideration of a flexible greenfield infrastructure build problem that does not, for example, restrict infrastructure expansion options to pre-defined paths or represent branches as binary integer variables. The utility in our approach is for the discovery of more general principles describing the optimal geographic scope of network design.

\section{Optimal Infrastructure Network Design}

Consider a system planner 
who has the task of designing (or modifying) an infrastructure system 
to provide a particular infrastructure product (water, natural gas, electricity, etc.) 
for a set of a set of $n$ locations (towns, buildings, etc.).
Each location $i$ has some known demand, $d_i$, for the product and also has the ability to produce this product locally with an incremental production cost, $c_i$, that varies with geography. 
In order to satisfy the demand at node $i$ one can either 
produce locally at a cost of $c_i$ per unit, 
or build an interconnection to some nearby node, 
with the intention of satisfying $d_i$ at a cost that is less than $c_i$.

Given a model of this sort we can ask a number of important questions.
Under what conditions is it optimal to produce locally, rather than building interconnections?
If the goal of the planner is to minimize overall cost, what type of network would one want to build? 
Should one build many small networks or one large one? 
Should the planner build a meshed network that allows redundancy, 
 or a radial network that provides only one path between sources and sinks?
 
In this section we introduce two relatively simple optimization models that allow one to address questions of this sort.
Both models use a ``greenfield'' approach, in which we seek to find the optimal network configuration that satisfies the total demand for a particular infrastructure product, given a set of objectives and constraints.

\subsection{Basic model}

To start, we assume that each of the $n$ locations is a vertex ($v\in V$) that has coordinates $x_v,y_v$ in some 2d space,
demand $d_v$, 
a per unit production cost of $c_v$ and
a maximum potential production capacity $\overline{g_v}$.
In addition we assume that there is a maximum feasible set of 
undirected edges $e \in E$ 
that one might choose to build.
For example one might allow into $E$ all possible vertex pairs, thus allowing at most 
$n(n-1)$ edges.
The cost of building any one particular edge 
$e_{i \leftrightarrow j}$ 
depends on two factors: 
the length of the edge 
\begin{equation}
    l_{e_{i \leftrightarrow j}} = \sqrt{(x_i-x_j)^2 + (y_i-y_j)^2}
\end{equation}
and the cost of one unit length and one unit capacity of edge construction, $w$.

Given these input data, the following formulation allows one to compute an ``optimal'' infrastructure network design.
\begin{subequations} \label{simple}
\begin{align}
    \min_{\overline{\mathbf{f}},\mathbf{f},\mathbf{g}} \quad   
                    & \sum_{v\in V}c_{v}g_{v} + 
                    w\sqrt{n}\sum_{e\in E} \ell_{e} \overline{f_{e}} \label{obj} \\
    \mathrm{s.t.} \quad     & 0\leq g_{v}\leq\overline{g_v},\forall v \label{g_lim} \\
                    & \overline{f_{e}} \ge 0, \forall e \label{f_cap_lim}\\
                    & -\overline{f_{e}}\leq f_{e}\leq\overline{f_{e}}, \forall e \label{f_lim} \\
                    & \mathbf{g} - \mathbf{d} = \mathbf{E}^{\intercal}\mathbf{f} \label{KCL}
\end{align}
\end{subequations}
where $\overline{f_e}$ and $f_e$ are the undirected flow capacity and actual directed flow on edge $e$,
$g_v$ is the actual amount of production at vertex $v$,
$w\sqrt{n}$ is an interconnection cost parameter (cost per-unit length$\cdot$capacity),
and 
$\mathbf{E}$ is an $m \times n$ edge matrix with 1 and -1 on the rows indicating the two endpoints for each of $m$ edges in the maximum feasible network.

Our objective (\ref{obj}) is to minimize the combined 
cost of production  $\mathbf{c}^\intercal \mathbf{g}$ and 
interconnection $\mathbf{\ell}^\intercal \overline{\mathbf{f}}$, 
while satisfying constraints (\ref{g_lim})-(\ref{KCL}).
Constraint (\ref{g_lim}) defines locational production limits;
(\ref{f_cap_lim}) ensures that we do not build negative quantities of interconnection capacity;
(\ref{f_lim}) constrains flows to be less than the chosen flow limits; 
and (\ref{KCL}) ensures that the net flow into and out of each vertex must be zero.

It is important to note that the cost function (\ref{obj}) is designed so that both the production and the edge construction cost terms grow linearly with $n$.
In order to implement this, we first observe that  
(at least for the case of uniformly distributed node locations, see Sec.~\ref{uniform1}) 
edge lengths $\ell_e$ fall with $n$ according to:
$\ell_e \sim n^{-1/2}$.
As a result, ensuring linear growth of the edge cost term requires that we multiply by $\sqrt{n}$, thus producing the term $w' = w\sqrt{n}$.

Implied in this formulation are a number of important assumptions.
First, we assume that interconnections can be built at any size scale and that construction costs scale linearly with the capacity of the edge. 
It is certainly possible to think of particular examples, such as transmission line construction, where costs are ``lumpy,''
such that building a 1 MW transmission line is more than 1/100 of the cost of building a 100 MW transmission line. 
However, if we consider that the edge might be either a large transmission line or a small distribution line, 
this assumption is not as obviously incorrect.
And modeling lumpiness of this sort would require knowledge about the details of a particular infrastructure system at a particular place and time. 
In this paper we are more interested to identify general trends that appear in optimal infrastructure designs.
Second, formulation (\ref{simple}) models a single snapshot of demand, whereas all real infrastructure systems have demand that varies in time. 
As a result the production cost term $c_v$ and the interconnection cost term $w$ meld together the capital and operating costs associated with supplying the demand $d_v$.
Finally, we make the assumption that nodes do not include any storage, 
leading us to include constraint (\ref{KCL}), 
which is equivalent to Kirchhoff's Current Law. 
Since our model aims to represent the long-term average operating pattern of a network,
rather than short-term time-domain details, 
we argue that this assumption provides at least some insight, 
even for networks that do include some storage in them, such as water networks.
Clearly, if one were wanting to design a particular infrastructure at a particular place and time, one would want to relax these assumptions and model additional details. 
However, this paper focuses on general trends that appear in optimal network designs.

\begin{figure*}[ht]
\includegraphics[width=1.0\textwidth]{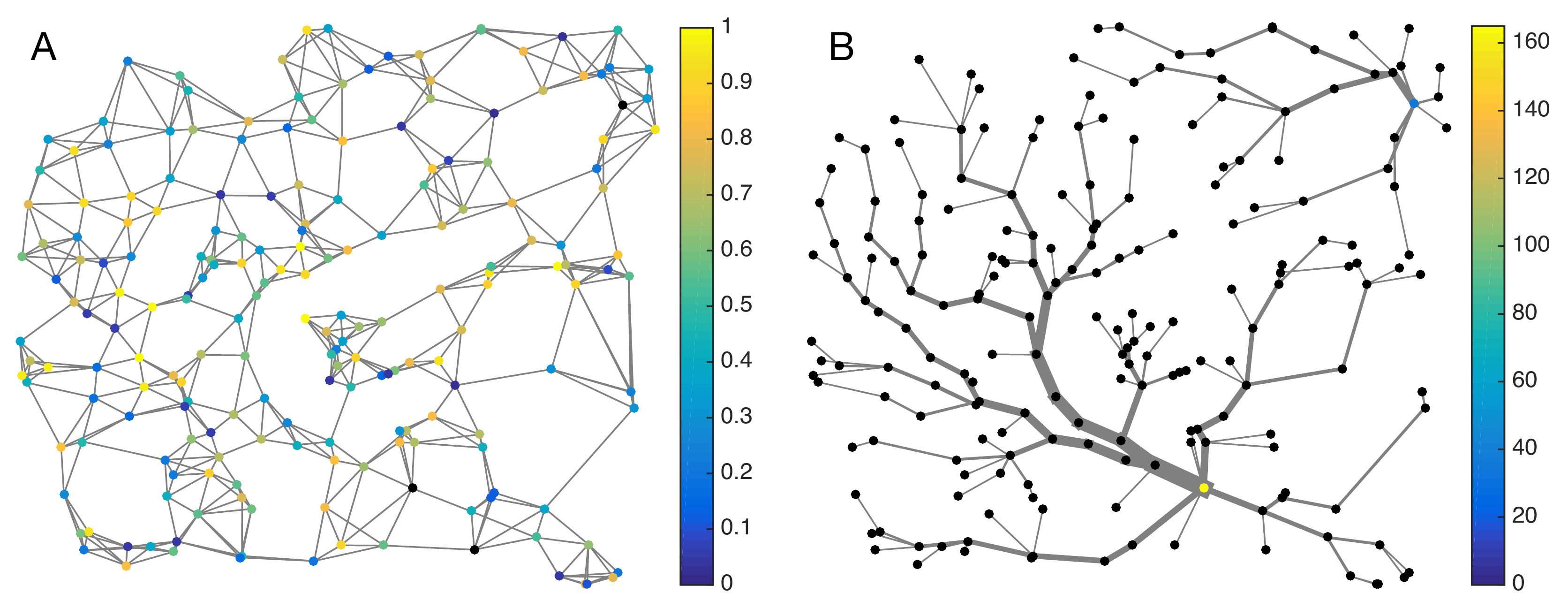}
\caption{
Illustrative results for random placement of ``towns'' on a 2d plane.
Panel A shows an initial, maximum feasible graph for a $n=200$ network, with node colors indicating the production cost, $c_v$, at each ``town.''
Panel B shows the optimal network configuration, 
after solving (\ref{obj})-(\ref{KCL}) for $w=0.001$, 
with node colors indicating the amount of production at each node, $g_v$, 
and edge thicknesses indicating the flow capacity.
This optimal network has two connected components and thus an average component size of $<n_s>=n/2=100$.
}
\label{fig:simple_illustration}
\end{figure*}

\subsection{Considering reliability}

An obvious limitation in the formulation above is the complete disregard of reliability. 
In reality, reliability has an enormous role in the design of infrastructure systems. 
In a power system, for example, 
electric utilities frequently argue (in their rate-case filings) that the construction of a new transmission line is justified purely on reliability grounds.
It is thus useful to understand what impact reliability considerations have on the topological structure of infrastructure networks.

In order to model the impact of reliability we add a third term to our objective function (\ref{obj}) to capture the cost of (un)reliability.
In addition, we separate the production cost term to include separate terms for capital and operating costs, 
since 
one will sometimes want to build surplus production capacity to prepare for plausible component failures.
With these additions to the objective and the associated constraints we get the following formulation.
\begin{subequations}\label{eq:reliability}
\begin{align}
\min_{f,\overline{f},g,\overline{g}} \quad & 
    \sum_{v\in V}\left(c_v g_{v} + k_v \overline{g_{v}}\right)
    +
    w\sqrt{n}\sum_{e\in E}\overline{f_{e}}l_{e}
    \nonumber
    \\
    & \qquad \qquad \qquad \qquad 
    + \frac{r}{n} \sum_{p\in P}\mathbf{1}^{\intercal}\Delta\mathbf{d}(p)
    \label{eq:cost}\\
\mathrm{s.t.} \quad
    & 
    0\leq g_{v}\leq\overline{g_{v}},\,\forall v
    \label{eq:gen_capacity}\\
    & 
    0\leq|f_{e}|\leq\overline{f_{e}},\,\forall e
    \label{eq:flow_limits}\\
    & 
    \mathbf{g}-\mathbf{d}=\mathbf{E}^{\intercal}\mathbf{f}
    \label{eq:KCL}\\
    & 
    \Delta g_{v}(p_{v})=-g_{v}, 
        \,\forall v\in\{1\ldots n\}
    \label{eq:node_failures}\\
    & 
    \Delta f_{e}(p_{e})=-f_{e}, 
        \,\forall e\in\{1\ldots m\}
    \label{eq:edge_failures}\\
    & 
    |f_{e}+\Delta f_{e}(p)|\leq\overline{f_{e}},
        \,\forall p\in P,\,\forall e\in E
    \label{eq:flows_after}\\
    & 
    0\leq g_{v}+\Delta g_{v}(p) \leq \overline{g_{v}},
        \,\forall p\in P,\,\forall v\in V
    \label{eq:gen_after}\\
    & 
    d_{v} \leq\Delta d_{v}(p)\leq 0,
        \,\forall p\in P,\,\forall v\in V
    \label{eq:demand_after}\\
    & 
    \left(\mathbf{g}+\Delta\mathbf{g}(p)\right)
    -\left(\mathbf{d}+\Delta\mathbf{d}(p)\right)
    \nonumber \\ 
    & \qquad \qquad \qquad \qquad
    = \mathbf{E}^{\intercal}\left(\mathbf{f}+\Delta\mathbf{f}(p)\right)
    \label{eq:KCL_after}
\end{align}
\end{subequations}
In this formulation,
(\ref{eq:cost}) is the modified objective, which now includes the reliability term.
In this term, $r$ is the reliability cost parameter that allows us to adjust the relative importance of reliability and
$\Delta \mathbf{d}(p)$ is the change (loss) of demand that results from perturbation $p$, which is one of the set of all perturbations, $P$. 
Eqns.~(\ref{eq:gen_capacity}) and (\ref{eq:KCL}) are equivalent to (\ref{g_lim})-(\ref{KCL}) in (\ref{simple}).
Eqs.~(\ref{eq:node_failures}) and (\ref{eq:edge_failures}) cause specific node, $p_v$, and edge, $p_e$, failures that together make up the set of all perturbations, $P$, by forcing the production or flow to be zero for the appropriate edge/perturbation combination.
While this approach could be used to model many different types of failures, here we consider $P$ to be the set of all single component (either production unit or edge) outages.
Eq.~(\ref{eq:flows_after}) ensures that all flows are below edge capacities, after all perturbations. As a result there are $m(n+m)$ constraints of this type within the formulation.
Similarly, (\ref{eq:gen_after}) constrains production at every node after each perturbation (a total of $n(m+n)$ constraints), to be below the chosen production capacities for each node.
Eq.~(\ref{eq:demand_after}) ensures that demand can only decrease, and only up to the total demand at node $v$, as a result of each $p$.
Finally, (\ref{eq:KCL_after}) enforces a nodal supply/demand balance after each perturbation. This forces the formulation to compute production, demand loss, and flow patterns that obey Kirchhoff's Current Law for each disturbance in $P$.

As a whole this formulation allows us to observe how network size and structure changes as we increase the relative importance of reliability.
If $r=0$, demand losses are effectively deemed irrelevant, and the problem will produce results that essentially identical to those obtained from \ref{simple}.
On the other hand, as $r$ increases we hypothesize that networks are likely to become more meshed (rather than tree-like) and more likely to include surplus production capacity.
It is not obvious, \emph{ex ante}, how $r$ will impact optimal network sizes.
On the one hand, small, local networks will be more robust to edge failures and thus may be more optimal when reliability is very important. 
On the other hand, large interconnected systems provide a high level of redundancy, which also has tremendous value.
In the sections that follow we explore this tradeoff.

\section{Results}

Here we explore the structural properties of the networks that emerge from formulations (\ref{simple}) and (\ref{eq:reliability}), under a variety of different cost and reliability assumptions.
Section \ref{uniform1} explores the case of nodes distributed uniformly within a 2d square, ignoring reliability cost.
Section~\ref{uniform_w_reliability} extends this to the reliability case and Sec.~\ref{senegal} applies our approach to data from the country of Senegal.

\subsection{Uniform distribution of load, ignoring reliability}\label{uniform1}

Consider the case of $n$ nodes randomly located within a $1 \times 1$ 2d square,
such that each node location $x_v$ and $y_v$ is a uniform random variable in $[0,1]$.
Each of these nodes has a production cost $c_v$ that is also a uniform random variable in $[0,1]$ and a demand $d_v = 1$.
The set of feasible edges that we might decide to build (the feasible graph $E$) comes from initially setting $E$ to be a modified form of the random geometric graph~\cite{dall2002random}.
In this case, rather than connecting each node to nodes that lie within a fixed radius, for each node $i$ an edge is added to connect from $i$ to $i$'s $k$ nearest (Euclidian) neighbors, while avoiding the addition of duplicate edges.
Because it is possible that $j$ is one of $i$'s $k$ nearest neighbors, 
but $i$ is not one of $j$'s $k$ nearest neighbors, 
the resulting $E$ has an average degree that is slightly larger than $k$. 
If one were to set $k \geq n-1$ the result would be the full graph of $n(n-1)/2$ possible edges.
However, optimizing over the full graph makes the resulting problem computationally impractical for all but the smallest problems.
Instead the results in this section come from the (somewhat arbitrary) choice of $k=5$.

Figure~\ref{fig:simple_illustration} illustrates the application of this approach to a system with $n=200$ nodes and $w=10^{-3}$.
From this figure a few observation can be made. 
First, we see that the algorithm tends to produce tree-like graphs 
in which the number of edges in each connected component, $m_s$, 
is one less than the number of nodes in that component $n_s$.
The reason for this is fairly straightforward:
creating a loop means that there are redundant paths between node pairs. 
Given a network with a loop, one can always reduce the edge-construction cost term 
$w\sqrt{n}\sum_{e\in E} \ell_{e} \overline{f_{e}}$ by removing one edge in the loop, without loss of functionality. 
As a result the graphs that result from (\ref{simple}) are always treelike, with precisely $m_s=n_s-1$ edges in each component.
Secondly, we see that there are two connected components in the optimal network 
and thus an average of $<n_s>=n/2=100$ nodes per component.
In this illustration the two least-expensive production nodes had costs of $c_{v_1}=0.0003$ and $c_{v_2}=0.0046$, with the less expensive node supplying the larger sub-component.
While it would have been feasible to connect the two components with a fairly short additional edge, supplying the whole network from the less expensive node $v_1$ would have required building additional capacity along the spidery path from $v_1$ to $v_2$.
Doing so would have cost more than the additional cost of supplying the second component from the more expensive unit, a cost of $n_{s_2}(c_{v_2}-c_{v_1})=0.149$.
For comparison purposes, the cost of building a length $\ell_e=1$ edge that could supply the whole of the 35 node second component would be $35w\sqrt{n}=0.495$.

Given that this approach can determine ``optimal'' network sizes, it is natural to ask how those network sizes change as the cost of building network infrastructure changes. 
For example, as $w$ increases one might expect to see a relatively sudden phase transition from optimal networks that span the entire network to optimal networks with many small, decentralized sub-systems.
In order to investigate this further and understand the cost conditions under which centralized, or decentralized, networks are optimal, 
we performed the following experiment.
For several values of $n$, we computed optimal infrastructure networks using (\ref{simple}) over a range of $w$ from $10^{-4}$ to 1.
For each value of $w$ we re-initialized the random node locations, 
the feasible network $E$ and production costs $c_v$, 
and computed the optimal network configuration using (\ref{simple}) 200 times. 
Then we recorded the mean size of the connected components $<n_s>$ over the 200 optimal networks.

\begin{figure}[H]
\centering
\includegraphics[width=1\columnwidth]{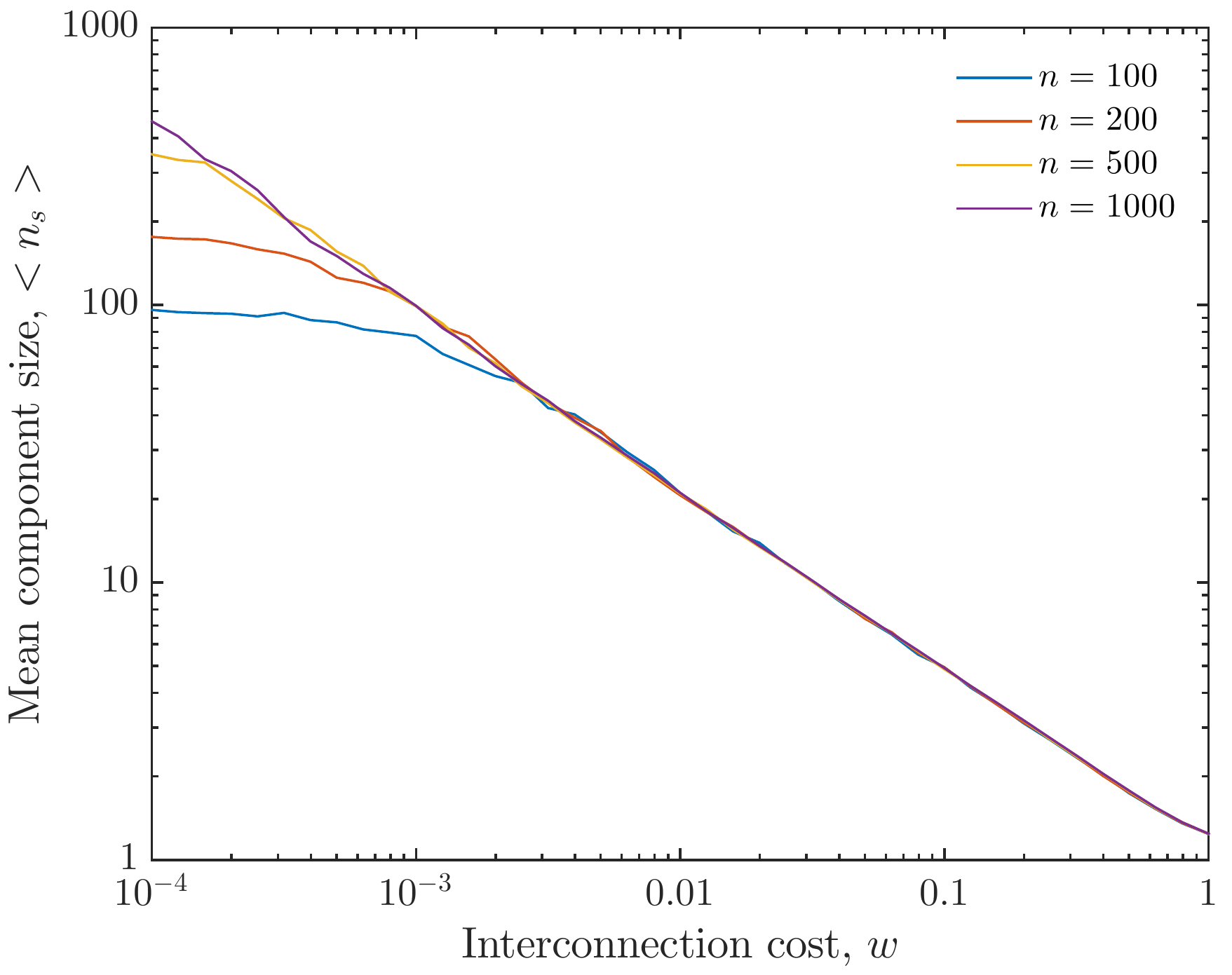}
\caption{Mean connected component sizes as a function of the interconnection cost parameter $w$. For each of the four different network sizes considered, we find that component sizes scale approximately as $w^{-2/3}$.}
\label{fig:simple_sizes}
\end{figure}

Figure \ref{fig:simple_sizes} shows the resulting relationship between the edge construction cost, $w$ and optimal component sizes.
As one would expect, 
as network construction costs increase the size of the optimal network decreases. 
However, what is somewhat surprising is that the change from large networks to small networks does not occur suddenly as does the first-order phase transition from a solid ice to liquid water. 
Instead, this transition occurs gradually over several orders of magnitude in $w$.
In fact, fitting the data in Figure~\ref{fig:simple_sizes} to a power-law distribution indicates that mean component sizes fall as
\begin{align}
  <n_{s}> \sim w^{-0.648} \sim w^{-2/3}. \label{eq:scaling}
\end{align}

\begin{figure*}[ht]
\centering
\includegraphics[width=1\textwidth]{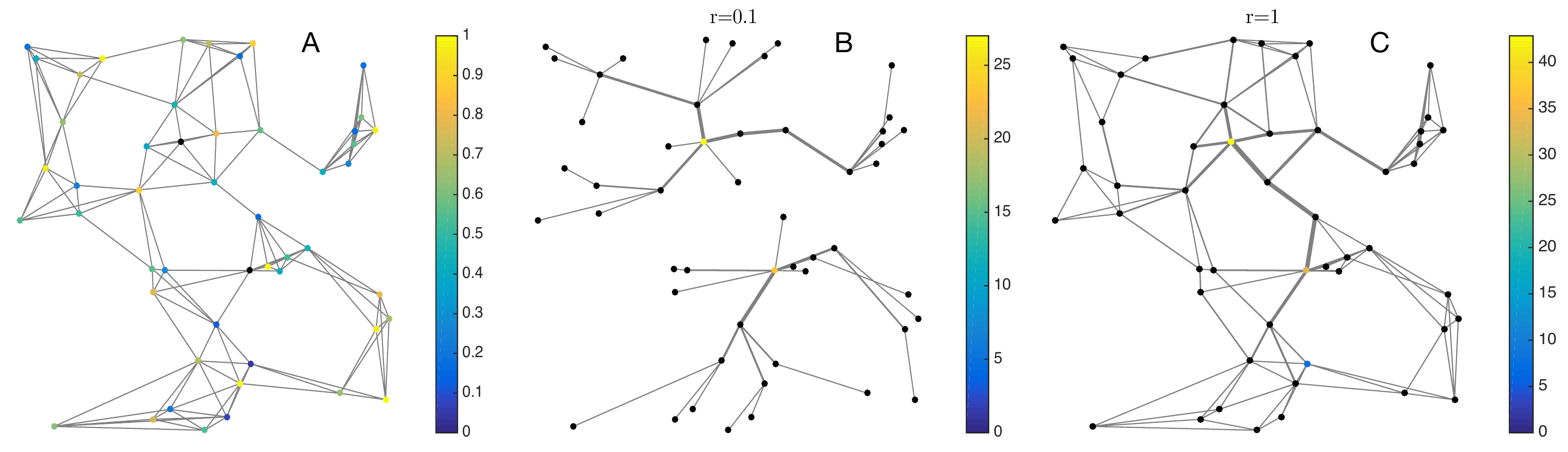}
\caption{Illustrating the impact of adding reliability to the optimal network construction formulation. 
Panel A shows the full feasible network $E$, with colors indicating production costs $c_v$. 
Panel B shows the optimal network for $w=0.01$ and $r=0.1$, which is  identical to the tree-like network that results from the simple model. 
Panel C shows the optimal network for $w=0.01$ and $r=1$, which shows the emergence of a meshed topology and substantial supply redundancy.
Colors in panels B and C indicate the amount of production capacity at each node. 
}
\label{fig:reliability_illustration}
\end{figure*}

\subsection{Uniform distribution of load, with reliability}\label{uniform_w_reliability}

Next we turn our attention to understanding how the results described above change after modeling reliability costs as in~(\ref{eq:reliability}).
As in the simple model, we consider nodes scattered uniformly on a 2d plane.
Also as before, we assume that each location has an overall production cost that is a uniform random variable in $[0,1]$, however unlike in the simple model we assume that this cost is split evenly between marginal and capital costs, $c_v$ and $k_v$.
Here we restrict our attention to the case of networks with $n=100$ nodes, since solving~(\ref{eq:reliability}) for larger systems leads to prohibitively large solution times.

First we show a few illustrative results for a $n=50$ node network
(see Fig.~\ref{fig:reliability_illustration})
that clearly show the importance of reliability to network structure. 
For small values of $r$ the solutions are nearly identical to what we get from the simple model: tree-like network that satisfy demand with no redundancy.
However as $r$ increases, we find a (rather sudden) transition to meshed networks that include substantial supply redundancy. 
For the example in Fig.~\ref{fig:reliability_illustration}, the system builds a network with total generation capacity equal to 
$\sum_{v\in V}\overline{g_v}=82.2$, much more than what is needed to supply the 50 nodes in the system.

\begin{figure}
\centering
\includegraphics[width=1\columnwidth]{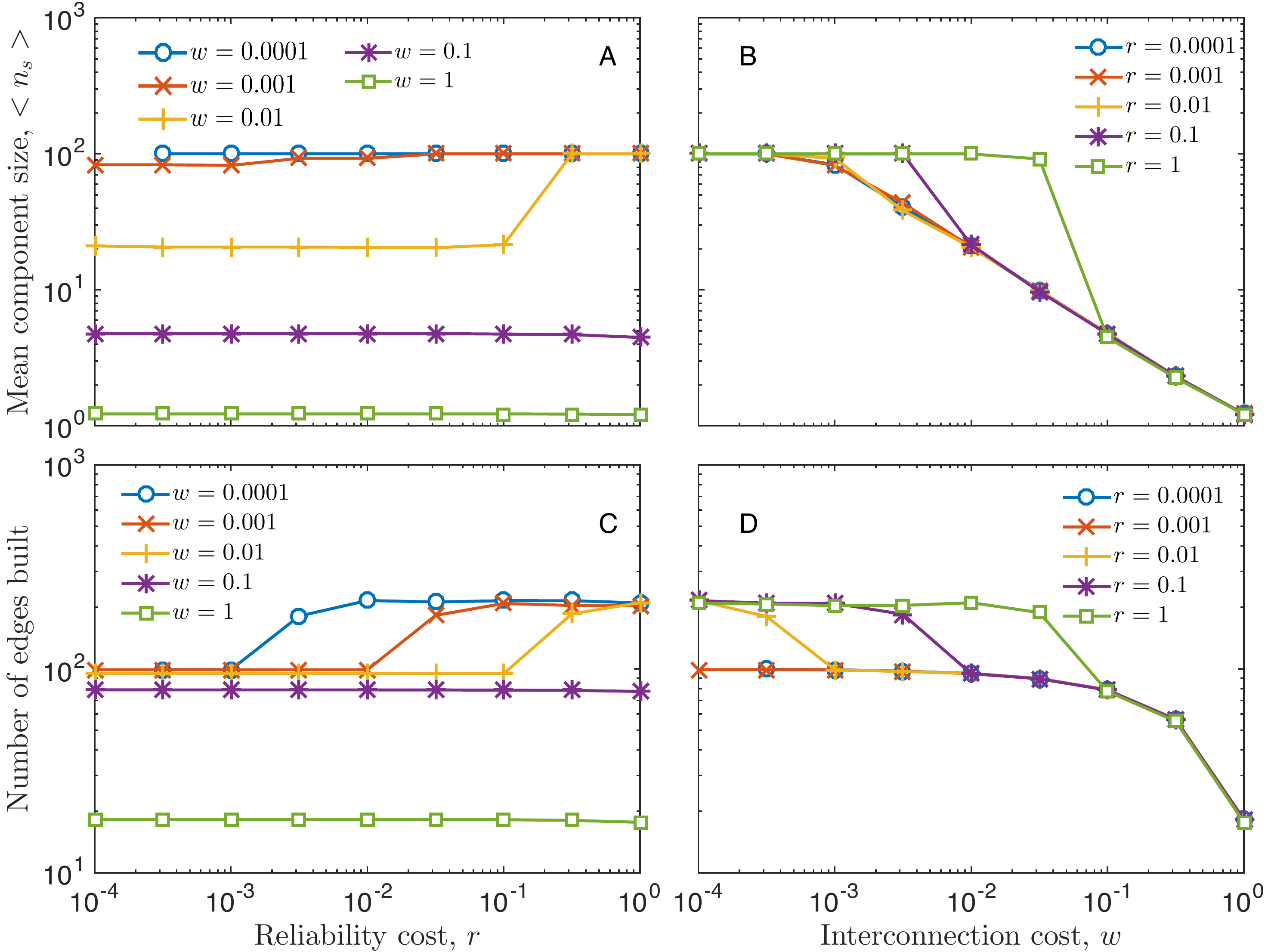}
\caption{Statistical results from the reliability model. 
Panels A and B show mean component sizes 
as a function of the reliability cost and interconnection cost parameters, $r$ and $w$.
Panels C and D show the average number of edges in the optimal network, 
also as a function of  $r$ and $w$.}
\label{fig:reliability_results}
\end{figure}

Next we computed optimal networks for several different values of the reliability parameter $r$ and interconnection costs $w$ for 100 nodes networks
For each value of $w$ and $r$ the random variables ($x_v,y_v$ and $c_v$) were re-initialized 100 times in order to minimize variance.%
\footnote{A few of these cases failed to solve, which means that a few of the results are averaged over fewer than 100 trials}
Figure~\ref{fig:reliability_results} shows the resulting mean component sizes for various values of $w$ and $r$.
For $w \leq 0.001$ and $w \geq 0.1$ the network sizes do not change substantially with $r$.
However for the intermediate case $w=0.01$ we see a sudden jump in optimal network size as $r$ passes 0.01.
For small $r$ the optimal size is around 20 nodes, whereas as reliability becomes more important the optimal network size increases toward the size of the network.
We also see a more sensitive relationship between component sizes and $w$ as $r$ increases. 
For example, Fig.~\ref{fig:reliability_results}B shows that for $r=1$ optimal network sizes suddenly decrease from full networks $<n_s>=n$ to relatively small ones $<n_s>\cong 4$, as $w$ increases above $0.03$.

Not only do the optimal network sizes change, but the level of redundancy also changes with $r$ and $w$.
One way to measure the level of redundancy is by the number of edges constructed in the optimal network. 
In the tree-like networks that result from the simple model there are always fewer than $n$ edges. 
But, as shown in Fig.~\ref{fig:reliability_results}, 
as $r$ increases the number of edges in the optimal networks also increase, frequently quite suddenly.

\subsection{Senegal application}\label{senegal}

\begin{figure*}[ht]
\centering
\includegraphics[width=1\textwidth]{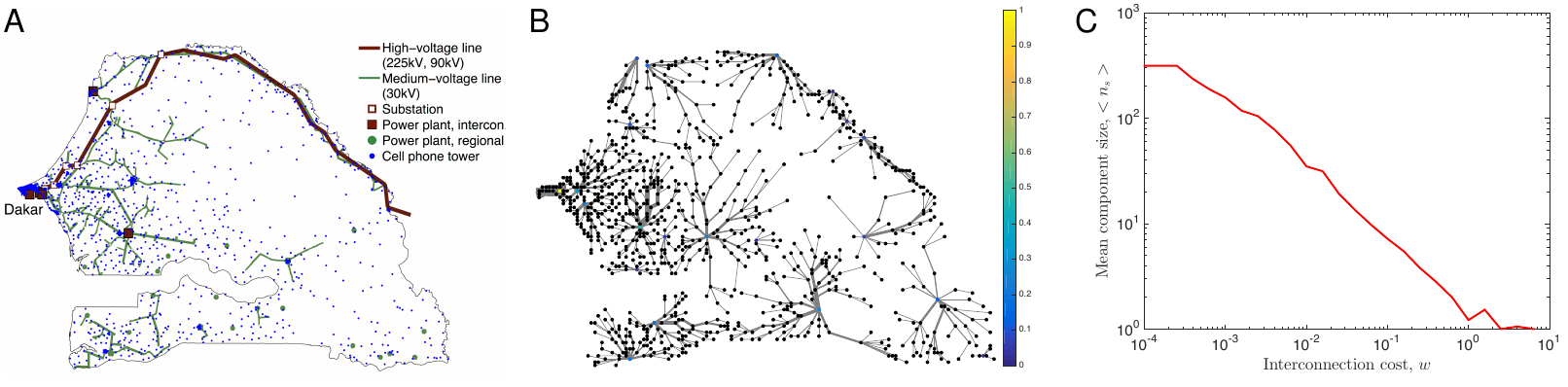}
\caption{Application of the optimal network design model to the geographic distribution of electricity demand centers in Senegal. Panel A shows the existing high and medium voltage network together with the geographic distribution of the mobile phone towers (adopted from~\cite{Cesena:2015}). Panel B shows the optimal network for $w=0.01$, given random production costs and ignoring reliability. Panel C shows the mean component size versus interconnection cost, $w$.}
\label{fig:senegal_results}
\end{figure*}

As a real-world case study, we applied the infrastructure network design model to the geographic distribution of cities and rural towns in Senegal. About half of the countries' population still has no access to electricity, and the electrification rate in rural areas is as low as 28$\%$~\cite{Cesena:2015}. 
In contrast to Senegal's electric power grid, the mobile communication infrastructure is highly developed, with 1666 mobile phone towers distributed across the country and a mobile phone penetration rate of almost 100$\%$. 
This allows for the use of data from the mobile communication system as a robust prediction for the geographic distribution of electricity needs, see~\cite{Cesena:2015} for technical details. Data on the mobile phone infrastructure has been made available by ORANGE / SONATEL within the framework of the D4D Challenge~\cite{D4D:2015}. Figure~\ref{fig:senegal_results}A depicts both the existing electricity infrastructure and the location of the mobile phone towers.  

In order to use data from the communication system to model demand for electricity, we first partitioned the country into a rectangular grid with cell size 5km $\times$ 5km. Following the approach in~\cite{Cesena:2015} we then used the number of cell phone towers that are located in each grid cell as a proxy for the relative electricity demand within that cell. Note that for the purpose of our analysis we are not interested in estimating absolute demand, but rather the relative amount of electricity that might be consumed in a particular location. 
The center points of the grid cells were used as locations $x_v,y_v$ for the load nodes.  
As with the uniformly distributed vertices, we randomly assigned production costs to each node in the network, using the load locations described above, using uniform random variables over $[0,1]$. 

Figure~\ref{fig:senegal_results}B shows the result of applying the basic optimization model (\ref{simple}) for the case of $w=0.01$. 
Interestingly, without reliability constraints, our model produces tree-like networks that 
i) are similar in structure to what we found with randomly distributed vertices in Sec~\ref{uniform1} and 
ii) closely resemble the tree-like topology of the existing electricity grid in Senegal. 
Moreover, we find that the scaling of optimal component sizes clearly follows the behavior previously (\ref{eq:scaling}) for synthetically generated load points (see Fig.~\ref{fig:senegal_results}C). 
This suggests that the power law decay in optimal network sizes is largely robust to changes in the geographic distribution of load locations.

\section{Conclusions}

This paper presents results from a model of optimal infrastructure network design for electric power production and delivery, with which we aimed to better understand the conditions under which electricity production and delivery were best managed through a highly connected network with large geographic reach, versus more localized networks with less global connectivity.
Several interesting observations emerged.
First we find, unsurprisingly, that as network construction costs increase the optimal size of infrastructure networks decreases and the local provision of electrical services becomes preferable. 
In the case where cost is the primary network design objective and reliability is not important, we unsurprisingly find that optimal topologies always have tree structure. 
More surprising, however, is the decrease in optimal network size occurs gradually, 
over several orders of magnitude in our network cost parameter, $w$.
More specifically, we find that optimal network sizes decrease with the power-law
$\sim w^{-2/3}$.
This same scaling property appears both in the random graphs that we generated for simulation purposes and when we apply our infrastructure design model to a spatial distribution of demand centers taken from data from the country of Senegal.
This suggests that when cost is the most important design criterion there is no single optimal size for infrastructure networks, 
but rather that different sizes are likely to be optimal for different locations. The distinction between which type of network architecture (local or global) is ``better'' is not clear.

We do find that this gradual scaling becomes a more sudden transition once reliability is added to the network design objectives.
When the failure to supply demand after vertex or edge outages is deemed costly (large $r$),
the optimal network is a single interconnected system that spans the entire network for a wide range of values for infrastructure costs $w$, and then a small increase in $w$ causes the optimal network size to be small.
Also, as the importance of reliability increases, 
the optimal network topology transitions from being a tree, in which there are no duplicate paths, to a meshed system with substantial redundancy. 

While the model that we used to reach these conclusions is simple, 
the results have important implications that may yield insight into difficult global problems such as the expansion of infrastructure in less developed countries and the potential transition from large power networks to smaller microgrids.

\bibliographystyle{ieeetr}
\bibliography{decentralized}

\section*{Acknowledgement}
The authors gratefully acknowledge the hospitality of the Santa Fe Institute in Santa Fe, NM, USA, 
where Hines and Blumsack were sabbatical visitors in 2014-2015 and where much of this work was completed. The authors would also like to acknowledge Christa Brelsford, Luis Bettencourt, and participants at the Santa Fe Institute workshop, ``Reinventing the Grid'' for helpful suggestions and discussions.

\section*{Author biographies}

\begin{IEEEbiographynophoto}{Paul D.~H.~Hines} (S`96,M`07,SM`14)
received the Ph.D.~in Engineering and Public Policy from Carnegie Mellon University in 2007 and M.S.~(2001) and B.S.~(1997) degrees in Electrical Engineering from the University of Washington and Seattle Pacific University, respectively.

He is currently an Associate Professor in the School of Engineering, and the L. Richard Fisher professor of electrical engineering, at the University of Vermont. 
Formerly he worked at the U.S.~National Energy Technology Laboratory, the U.S.~Federal Energy Regulatory Commission, Alstom ESCA, and for Black and Veatch. He currently serves as the chair of the Green Mountain Section of the IEEE, as the vice-chair of the IEEE PES Working Group on Cascading Failure, and as an Associate Editor for the IEEE Transactions on Smart Grid. He is a National Science Foundation CAREER award winner.
\end{IEEEbiographynophoto}

\begin{IEEEbiographynophoto}{Seth Blumsack} (M `06) received the Ph.D.~in Engineering and Public Policy from Carnegie Mellon University in 2006, the M.S. degree in Economics from Carnegie Mellon in 2003 and the B.A. degree in Mathematics and Economics from Reed College in 1998. He is currently an Associate Professor in the John and Willie Leone Family Department of Energy and Mineral Engineering at The Pennsylvania State University, Chair of the Energy Business and Finance program and is the John T. Ryan, Jr. Fellow in the College of Earth and Mineral Sciences.
\end{IEEEbiographynophoto}

\begin{IEEEbiographynophoto}{Markus~Schl\"apfer} received the Ph.D.~in Mechanical and Process Engineering~(2010) and the M.S. in Environmental Engineering~(2003), both from ETH Zurich. He is currently a Postdoctoral Fellow at the Santa Fe Institute and a Research Affiliate at MIT's Senseable City Lab. 
\end{IEEEbiographynophoto}

\end{document}